\documentclass[a4paper,twoside,british]{elsarticle}
\usepackage[T1]{fontenc}
\usepackage[latin9]{inputenc}
\usepackage{units}
\usepackage{amsmath}
\usepackage{graphicx}
\usepackage{setspace}
\makeatletter

\journal{ }

\usepackage{lineno}

\makeatother

\usepackage{babel}
\begin{document}


\title{Comparative study on mesh-based and mesh-less coupled CFD-DEM methods to model particle-laden flow}

\author[RUB]{D.~Markauskas\corref{cor1}}

\ead{markauskas@leat.rub.de, Tel: +49 (0)234 / 32-27281, Fax: +49 (0)234 / 32-14227 }

\author[RUB,TUB]{H.~Kruggel-Emden}

\author[US]{R.~Sivanesapillai}

\author[US]{H.~Steeb}

\cortext[cor1]{Darius Markauskas}

\address[RUB]{Ruhr University Bochum, Universit\"atsstrasse 150, D-44780 Bochum,
Germany}

\address[TUB]{Technical University of Berlin, Ernst-Reuter Platz 1, D-10587 Berlin,
Germany}

\address[US]{University of Stuttgart, Pfaffenwaldring 7, D-70569 Stuttgart, Germany}
\begin{abstract}
A comparative study on mesh-based and mesh-less Computational
Fluid Dynamics (CFD) approaches coupled with the Discrete Element Method (DEM)
is presented. As the mesh-based CFD approach a Finite Volume Method
(FVM) is used. A Smoothed Particle Hydrodynamics (SPH) method represents
mesh-less CFD. The unresolved fluid model is governed by the locally averaged
Navier-Stokes equations. A newly developed model for applying boundary conditions in the SPH
is described and validation tests are performed. With the help of the presented comparative tests,
the similarities and differences of DEM-FVM and DEM-SPH methods are discussed. 
Three test cases, comprised of a single solid particle sedimentation test, flow through a
porous block and sedimentation of a porous block, are performed using both methods.
Drag forces acting on solid particles highly depend on local fluid fractions. For comparative reasons,
the size of a cell in FVM is chosen such that fluid fractions match those computed in SPH. In
general, DEM-FVM and DEM-SPH methods exhibit good agreement with analytic reference results.
Differences between DEM-SPH and DEM-FVM approaches were found mostly due to differences in computed local fluid fractions.
\end{abstract}
\maketitle


\section{Introduction}

Coupled particle-fluid flow can be observed in almost all types of
particulate processes. Existing approaches to model particle-fluid
flow can be classified into two categories {[}\citealp{Hoef2004}{]}:
the discrete approach at microscopic level (particles are resolved
as separate bodies) and the continuum approach at the macroscopic
level (the fluid phase as well as the particle phase are modeled as 
fully inter-penetrating). In the continuum approach, macroscopic
behavior is governed by balance equations (e.g. for mass and momentum)
closed with constitutive relations together with initial and boundary
conditions {[}\citealp{Gidaspow1994}{]}. This approach is preferred in
process modeling and applied research because of its computational efficiency.
However, its effective use heavily depends on the constitutive relations and
the model that accounts for momentum exchange between particles and 
fluid phase {[}\citealp{Zhu2007}{]}.

Discrete simulation approaches to model particle flow are based on
the analysis of the motion of individual particles, e.g. using the
Discrete Element Method (DEM), and thus inherit a reduced set of constitutive
assumptions as compared to continuum approach. In this approach,
the fluid phase can be modeled at the sub-particle level (Direct Numerical
Simulation, DNS) such that momentum exchange (fluid-particle,
particle-particle) is resolved in detail {[}\citealp{Han2007}{]}, or at the coarse-grained
level (unresolved simulations using local volume-average technique) used for larger scale models
{[}\citealp{Deen2007}{]}. The simulation at the sub-particle level
can be used, e.g. for a detailed analysis of interaction forces that act between
the fluid phase and suspended particles or for investigating the behavior of complex
shaped particles dispersed in the fluid. However, such simulations are
usually limited to a small number of particles {[}\citealp{Wang2010,Zhao2013}{]}.
The unresolved approach is computationally more efficient and
allows simulation of much larger particle systems than DNS, while
preserving discrete flow characteristics of the particles. 

In most unresolved simulations mesh-based Computational
Fluid Dynamics (CFD) methods are used. A Finite Volume Method (FVM) for the
gas phase based on the locally averaged Navier-Stokes equations and
DEM for the solid phase was first reported by Tsuji et al. {[}\citealp{Tsuji1993}{]}.
Since then, a lot of investigations for the improvement of various
aspects of this coupling were conducted {[}\citealp{Zhu2007,Deen2007}{]}.
A wide range of applications such as fluidized beds {[}\citealp{Kafui2011}{]},
cyclones {[}\citealp{Chen2012}{]}, screening {[}\citealp{Li2012}{]},
pipeline flow {[}\citealp{Li2014}{]}, particle coating processes
{[}\citealp{Darabi2011}{]}, pneumatic particle transport {[}\citealp{Emden2014}{]}
and others have been discussed in the scientific literature. It could
be concluded, that DEM techniques coupled with mesh-based methods
are widely recognized as state of the art in current research.

A different situation prevails when mesh-less methods are applied for
the fluid phase. Coupling of DEM with the mesh-less Smoothed Particle
Hydrodynamics (SPH) method was investigated only in a couple of contributions.
Potapov et al. {[}\citealp{Potapov2001}{]}, Qiu {[}\citealp{Qiu2013}{]},
Canelas et al. {[}\citealp{Canelas2016}{]}
presented a two-way coupled DEM-SPH method. Because a DNS approach
was used, the method is suitable for modeling of few solid particles
only. Li et al. {[}\citealp{Li2007}{]} developed a SPH model for pore fluid
flows through solid particle packings, however, the model does not allow
for an independent movement of the fluid and solid particles. Jiang
et al. {[}\citealp{Jiang2007}{]} used SPH for modeling fluid flow
in isotropic porous media, however, the solid particles representing the
porous media remained static throughout the simulations. The analyses
of slurry transport in SAG mills and large screens were presented
in {[}\citealp{Cleary2006}{]} and {[}\citealp{Fernandez2011}{]},
where fluid flow and solid particle motion were computed using SPH and DEM, respectively.
However, the model represented a one-way coupling
between DEM and SPH only. Recently, a two-way coupling scheme
between DEM and SPH has been derived by Gao and Herbst {[}\citealp{Gao2009}{]}, Sun et al. {[}\citealp{Sun2013}{]}
and Robinson et al. {[}\citealp{Robinson2014}{]}. The application
to slurry flow, abrasive wear and magnetorheological fluids were demonstrated by Cleary {[}\citealp{Cleary201585}{]}, Beck \& Eberhard {[}\citealp{Beck2015}{]} and 
Lagger et al. {[}\citealp{Lagger2015}{]} respectively. These
first results look promising, however more investigations are required
to clarify various aspects of DEM-SPH coupling.

In the current investigation a comparison between DEM coupled
with FVM (mesh-based) and DEM coupled with SPH (mesh-less)
is presented. Some of the effects influencing the motion of suspended solid
particles are highlighted and the similarities/differences of solid particle
motion in both methods are discussed. Sections \ref{sec:Governing-eq-solid},
\ref{sec:Governing-eq-fluid} and \ref{sec:Interaction} describe
the governing equations of the fluid and solid phases and the interaction
between them. A newly developed model for boundary conditions in the
SPH is described and validation tests are performed in section \ref{sec:Boundaries}.
Three test cases, comprised of a single particle sedimentation test, flow
through a porous block and sedimentation of a porous block, are performed
using DEM-FVM and DEM-SPH methods whose results are discussed in sections
\ref{sec:Single-part-test}, \ref{sec:gravity-block-test} and \ref{sec:Porous-block-test}.

\section{Governing equations of the solid phase\label{sec:Governing-eq-solid}}

The solid phase is modeled using DEM. In this method the motion of each individual solid particle $\mathcal{P}_{i}$ is described by Newton's second law:

\begin{equation}
m_{i}\frac{d\mathbf{u}_{i}}{dt}=\mathbf{F}_{i}^{c}+\mathbf{F}_{i}^{g}+\mathbf{F}_{i}^{int},\label{eq:sp_newton}
\end{equation}
where $\mathbf{u}_{i}$ denotes the solid particle velocity, $\mathbf{F}_{i}^{c}$
denotes the total contact force, $\mathbf{F}_{i}^{g}$ denotes the gravity force and
$\mathbf{F}_{i}^{int}$ denotes the interaction force between solid and
fluid phase. The calculation of $\mathbf{F}_{i}^{int}$ is described
later in the section \ref{sec:Interaction}. The total contact force for solid
particle $\mathcal{P}_{i}$ is obtained from the sum of contact forces acting 
between $\mathcal{P}_{i}$ and its neighboring solid particles $\mathcal{P}_{j}$ :

\begin{equation}
\mathbf{F}_{i}^{c}=\sum_{j=1}^{n}\mathbf{F}_{ij}^{c}\:,\label{eq:ContactSum}
\end{equation}
where $n$ denotes the number of contacts. The contact force between solid particles 
is calculated as a sum of normal and tangential force components. A linear 
spring damper model is used for the normal component of the contact force. 
A linear spring limited by the Coulomb condition is used for the tangential force. 
A more detailed description
of the used DEM model can be found in {[}\citealp{Emden2014,Markauskas2015}{]}.

Discontinuities, such as an instant
application of external forces, lead to spurious high-frequency oscillations
in weakly-compressible SPH methods. To reduce this artifact Adami
et al. {[}\citealp{Adami2012}{]} proposed to increase the external
force gradually. In our case, the proposed technique is used for gradual
increase of the gravity force acting on a solid particle:

\begin{equation}
\mathbf{F}_{i}^{g}=V_{i}\mathbf{g}\left[\rho_{f}+\left(\rho_{s}-\rho_{f}\right)\xi(t)\right]\label{eq:f_g_damping}
\end{equation}
where $V_{i}$ denotes the volume of the solid particle, $\mathbf{g}$
denotes gravitational acceleration, $\rho_{f}$ denotes fluid density, $\rho_{s}$ denotes the density
of the solid particle and $\xi$ is a damping factor {[}\citealp{Adami2012}{]}:

\begin{equation}
\xi(t)=0.5\left[\sin\left(\pi\left(-0.5+\frac{t}{t_{damp}}\right)\right)+1\right],\; t\leq t_{damp}.\label{eq:t_damp}
\end{equation}
where $t_{damp}$ is the predefined damping time during which the force
gradually increased until the nominal value is reached.

\section{Governing equations of the fluid phase\label{sec:Governing-eq-fluid}}

\subsection{Mesh-based model}

The local averaging technique {[}\citealp{Anderson67}{]} for the
Navier-Stokes equations is applied in this research. This technique
is used widely for modeling fluid-particle interaction when unresolved
particle-fluid flow is considered {[}\citealp{Deen2007,Zhu2007}{]}.
The fluid phase is described in an Eulerian framework where continuity
and momentum equations are given as

\begin{equation}
\frac{\partial\bar{\rho_{f}}}{\partial t}+\nabla\cdot(\bar{\rho}_{f}\mathbf{u}_{f})=0,\label{eq:continuity}
\end{equation}

\begin{equation}
\frac{\partial\bar{\rho}_{f}\mathbf{u}_{f}}{\partial t}+\nabla\cdot(\bar{\rho}_{f}\mathbf{u}_{f}\otimes\mathbf{u}_{f})=-\varepsilon\nabla p+\nabla\cdot(\varepsilon\boldsymbol{\tau})-\mathbf{f}_{m}^{int}+\bar{\rho}_{f}\mathbf{g}\textrm{,}\label{eq:fvm_momentum}
\end{equation}
where $\bar{\rho}_{f}=\varepsilon\rho_{f}$ denotes the superficial (locally
averaged) density of the fluid, $\varepsilon$ denotes the local mean fluid
volume fraction, $\mathbf{u}_{f}$ denotes fluid velocity, $p$ denotes
pressure, $\boldsymbol{\tau}$ denotes the viscous stress tensor and
$\mathbf{f}_{m}^{int}$ denotes the particle-fluid interaction force per
unit of volume. The interaction force is further introduced in section \ref{sec:Interaction}.
The required porosity $\varepsilon$ in each fluid cell is calculated as follows. Each fluid cell is divided into a number of smaller sub-cells, called a transfer grid in {[}\citealp{Bluhm-Drenhaus2010}{]}. During the calculation the occupation of each sub-cell is checked. If the sub-cell center is inside of a solid particle, the volume of the sub-cell is marked. From the number of not marked sub-cells the approximate part of volume not occupied by solid particles is calculated for each fluid cell. This part of volume is used in Eq. (\ref{eq:momentum}) as the porosity of the cell. In the current study each fluid cell was divided into 40x40x40 sub-cells.
The presented momentum equation Eq. (\ref{eq:momentum}) corresponds to the model A as described by
Feng and Yu {[}\citealp{Feng2004}{]}.

For the mesh-based model a Finite Volume Method (FVM) as implemented
in the commercial software ANSYS Fluent is applied. The  porous media single phase model is used where the fluid is assumed as incompressible. In the performed tests neither the solid particles nor the fluid are heated.  Ansys ICEM meshing software is used for the generation of meshes. 
The coupled analysis at each time step consists of DEM part and FVM part. The updated positions of solid particles from DEM are used for the calculation of the porosity and the interaction force. For the transfer of information about fluid velocities from the FVM and to assign the calculated porosity as well as the interaction force User Defined Functions of Fluent are utilized. More details about the used DEM-FVM coupling algorithm can be found in {[}\citealp{Bluhm-Drenhaus2010}{]}.

\subsection{Mesh-less model}

The Smoothed Particle Hydrodynamics (SPH) method is used as a mesh-less
CFD method as an alternative approach to be coupled to the DEM. SPH is a mesh-less Lagrangian
technique first introduced by Gingold and Monaghan {[}\citealp{Gingold1977}{]}
and Lucy {[}\citealp{Lucy1977}{]} to solve problems of gas dynamics
in astrophysics. Since then it has also found a widespread use in
other areas of science and engineering. Its mesh-less character
makes the method very flexible and enables the simulations of physical
problems that might be difficult to capture by conventional mesh-based
methods. The principal idea of SPH is to treat hydrodynamics in a
completely mesh-free fashion, in terms of a set of sampling particles
{[}\citealp{Monaghan2005}{]}. SPH particles represent a finite, lumped mass
of the discretized continuum and carry information about all physical
variables evaluated at their positions. Hydrodynamic equations for
motion are then derived for these particles thus yielding a quite
simple formulation of fluid dynamics. Mass and linear momentum
are simultaneously conserved. Function values
and their derivatives at a specific SPH particle are interpolated from
function values at surrounding SPH particles using the interpolating (kernel)
function and its derivatives. Because of the mesh-free nature of SPH,
it can easily deal with problems characterized by large displacements
of the fluid-structure interface, by a rapidly moving fluid free-surface
and by complicated geometric settings. SPH has been applied across
a broad range of engineering disciplines to compute various environmental
or industrial fluid flows, for example, in marine {[}\citealp{Shibata2007}{]},
extrusion {[}\citealp{Prakash2015}{]}, geophysical {[}\citealp{Sivanesapillai2014}{]}
and coastal {[}\citealp{Gotoh2006}{]} engineering.

In a Lagrangian framework the continuity equation and the momentum
equation following {[}\citealp{Robinson2014}{]} are used for the fluid phase:

\begin{equation}
\frac{D\bar{\rho_{f}}}{Dt}+\nabla\cdot(\bar{\rho}_{f}\mathbf{u}_{f})=0,\label{eq:continuity-Lag}
\end{equation}

\begin{equation}
\frac{D\bar{\rho}_{f}\mathbf{u}_{f}}{Dt}=-\nabla p+\nabla\cdot(\varepsilon\boldsymbol{\tau})-\mathbf{f}^{int}+\bar{\rho}_{f}\mathbf{g}\textrm{,}\label{eq:momentum}
\end{equation}
The Eq.(\ref{eq:momentum}) corresponds to model B in {[}\citealp{Feng2004,Zhu2007}{]},
in which is assumed that the pressure gradient is applied to the fluid
phase only.

In the SPH method, the fluid phase is represented by separate particles. These
particles carry variables such as velocity, pressure, density and mass. No
connectivity is modeled between fluid particles, however, the integral representation
of the function is approximated by summing up the values of the neighboring
points using smooth kernel functions.

As commonly used in the weakly compressible approach to simulate
incompressible fluids, an equation of state is introduced to estimate
the pressure from the density field {[}\citealp{Colagrossi2003,Monaghan2005}{]}:

\begin{equation}
p=\frac{\rho_{0}c^{2}}{\gamma}\left[\left(\frac{\bar{\rho_{f}}}{\varepsilon\rho_{0}}\right)^{\gamma}-1\right]+B,\label{eq:state}
\end{equation}
where $\rho_{0}$ denotes the initial density of the fluid phase and $c$
denotes the speed of sound. To keep the density to vary by at most 1\%
with respect to $\rho_{0}$,
$c=10u_{max}$ is usually used {[}\citealp{Morris1997,Colagrossi2003}{]},
where $u_{max}$ denotes the maximum fluid velocity magnitude. The coefficient
$\gamma=7$ is used in our simulations. $B$ denotes a background pressure,
which is set to zero in case of free surface problems, while $B>0$
is used to avoid the tensile instability in other cases {[}\citealp{Colagrossi2003,Adami2012,Marrone2013}{]}. 

The kernel function is defined so that its value monotonously decreases
as the distance between SPH particles increases. It has a compact support,
the radius of which is defined by the smoothing length. The
Gaussian {[}\citealp{Monaghan1983}{]}, quadratic {[}\citealp{Dalrymple2006}{]},
cubic {[}\citealp{Monaghan1999,Gomez2004}{]} and quintic spline {[}\citealp{Wendland1995}{]}
as well as other functions can be used for this purpose. In the current
research a cubic spline {[}\citealp{Monaghan1985}{]} is used as kernel
function:

\begin{equation}
W(r,h)=\alpha_{D}\begin{cases}
1-\frac{3}{2}q^{2}+\frac{3}{4}q^{3}, & \quad0\leq q<1,\\
\frac{1}{4}(2-q)^{3}, & \quad1\leq q<2,\\
0, & \quad q\geq2,
\end{cases}\label{eq:KernelSpline}
\end{equation}
where $q=r/h$, $\alpha_{D}$ is $10/(7\pi h^{2})$ in case of 2D,
while $1/(\pi h^{3})$ in the 3D case. The smoothing length,
which defines the influence area of the kernel, is denoted $h$. 
The distance between two fluid particles $\mathcal{P}_a$ and $\mathcal{P}_b$
is denoted $r_{ab}=\Arrowvert\mathbf{r}_a-\mathbf{r}_b\Arrowvert$.

The continuity equation (\ref{eq:continuity-Lag}) discretized using SPH 
and evaluated for a fluid particle $\mathcal{P}_a$ takes the form 

\begin{equation}
\frac{D\bar{\rho}_{a}}{Dt}=\underset{b}{\sum}m_{b}\mathbf{u}_{ab}\cdot\nabla_{a}W_{ab},\label{eq:continuity-SPH}
\end{equation}
where indexes $a$ and $b$ indicate variables evaluated at positions $\mathbf{r}_a$ and $\mathbf{r}_b$
of fluid particles $\mathcal{P}_a$ and $\mathcal{P}_b$, respectively. $m_b$ denotes the
mass of fluid particle $\mathcal{P}_b$. $\mathbf{u}_{ab}=\mathbf{u}_{a}-\mathbf{u}_{b}$ is the relative
velocity between fluid particles $\mathcal{P}_a$ and $\mathcal{P}_b$. $\nabla_{a}W_{ab}=\nabla_{a}W(r_{ab},h)$
is the gradient of the kernel function. The summation is performed
over all neighboring fluid particles (these are with index $b$) of fluid particle $\mathcal{P}_a$.

The momentum conservation equation (\ref{eq:momentum}) in SPH takes
the form {[}\citealp{Morris1997}{]}:

\begin{equation}
\begin{array}{c}
\frac{D\mathbf{u}_{a}}{Dt}=-\underset{b}{\sum}m_{b}\left(\frac{p_{a}}{\bar{\rho}_{a}^{2}}+\frac{p_{b}}{\bar{\rho}_{b}^{2}}\right)\nabla_{a}W_{ab}+\mathbf{g}\\
+\underset{b}{\sum}m_{b}\frac{\nu(\bar{\rho}_{a}+\bar{\rho}_{b})}{\bar{\rho}_{a}\bar{\rho}_{b}}\cdot\frac{\mathbf{r}_{ab}\nabla_{a}W_{ab}}{|\mathbf{r}_{ab}|^{2}+\delta^{2}}\mathbf{u}_{ab}+\frac{\mathbf{f}_{a}^{int}}{m_{a}},
\end{array}\label{eq:momentum-SPH}
\end{equation}
where $\mathbf{r}_{ab}=\mathbf{r}_a-\mathbf{r}_b$.
The third term on the right-hand side in Eq. (\ref{eq:momentum-SPH}) is
a viscous term introduced by Morris {[}\citealp{Morris1997}{]}, where
$\nu$ denotes the kinematic viscosity. $\delta$ is a small, positive number used
to keep the denominator non-zero and usually set to $0.1h$.
In case of solid particles approaching the fluid particle, the resulting porosity $\varepsilon$ is decreasing and this is the cause of increased pressure in the fluid calculated by Eq. (\ref{eq:continuity-SPH}). Increased pressure will cause increased forces between pairs of fluid particles (see the first right term in Eq. (\ref{eq:momentum-SPH})). In this way, the fluid particles are pushed away form the approaching solid particles.

$\mathbf{f}_{a}^{int}$ in Eq. (\ref{eq:momentum-SPH}) represents the solid-fluid
interaction force acting on the fluid particle $\mathcal{P}_a$ due to momentum exchange with solid
particles. The force $\mathbf{f}_{a}^{int}$ is calculated as the
sum over solid particles in the domain of the fluid particle $\mathcal{P}_a$:

\begin{equation}
\mathbf{f}_{a}^{int}=\underset{i}{\sum}\mathbf{f}_{ai}^{int}\:,
\end{equation}

\begin{equation}
\mathbf{f}_{ai}^{int}=-\frac{V_{a}W_{ai}}{\underset{b}{\sum}V_{b}W_{bi}}\mathbf{F}_{i}^{int}\:.\label{eq:f_int_S-F}
\end{equation}
where $V_{a}$ denotes the volume of the fluid particle  $\mathcal{P}_a$ calculated as $V_{a}=m_{a}/\bar{\rho}_{a}$, while $\mathbf{F}_{i}^{int}$
is the same interaction force as in Eq. (\ref{eq:sp_newton}).

The fluid volume fraction $\varepsilon_{a}$ at the position of the fluid particle $\mathcal{P}_a$ is calculated from
the volumes of all solid particles $\mathcal{P}_i$ which are in the smoothing domain of the fluid particle $\mathcal{P}_a$:

\begin{equation}
\varepsilon_{a}=1-\sum_{i}V_{i}W_{ai}\:,\label{eq:fluid-fraction}
\end{equation}
where $V_{i}$ denotes the volume of the solid particle $\mathcal{P}_i$, while $W_{ai}=W(\Arrowvert\mathbf{r}_a-\mathbf{r}_{i}\Arrowvert,h)$
is evaluated using Eq. (\ref{eq:KernelSpline}).

Special care should be taken during the calculation of the fluid fraction
for the fluid particles near boundaries. If the boundary intersects
the kernel domain, a part of the kernel domain is truncated by the 
domain boundary. This implies truncation errors in the computation of $\varepsilon$. To
account for kernel domain truncation, the fluid fraction near
boundaries is calculated by:

\begin{equation}
\varepsilon_{a}=1-\frac{\underset{i}{\sum}V_{i}W_{ai}}{\Gamma_{a}}\label{eq:fluid-fraction-boundary}
\end{equation}
where $\Gamma_{a}$ is a correction factor for $\mathcal{P}_a$
as is described by Sun et al. {[}\citealp{Sun2013}{]}. It is an integral over the part of the kernel function which is inside of the problem domain. This modification 
normalizes the interpolation scheme in the vicinity of boundaries to reduce truncation errors. Originally, 
the $\Gamma_{a}$ factor was proposed in {[}\citealp{Kulasegaram2004}{]}
for the developed boundary model and was later modified in {[}\citealp{Sun2013}{]}.
$\Gamma_{a}$ for the cubic spline kernel is calculated from:

\begin{equation}
\Gamma_{a}=\begin{cases}
-\frac{1}{60}\left(3\psi^{6}-9\psi^{5}+20\psi^{3}-42\psi-30\right), & 0<\psi\leq1,\\
\frac{1}{60}\left(\psi^{6}-9\psi^{5}+30\psi^{4}-40\psi^{3}+48\psi+28\right), & 1<\psi\leq2,\\
1, & \psi>2,
\end{cases}
\end{equation}
where $\psi=y/h$, $y$ denotes the normal distance between the rigid boundary
and the position of a fluid particle and $h$ is the smoothing length
as it is used in the kernel function Eq.(\ref{eq:KernelSpline}).

\section{Fluid-solid interaction\label{sec:Interaction}}

The interaction force $\mathbf{F}_{i}^{int}$ acting on solid particle $\mathcal{P}_i$
can consist from several individual solid-fluid interaction forces {[}\citealp{Zhu2007}{]}.
In the present study, the drag force $\mathbf{F}_{i}^{D}$ and the pressure
gradient force $\mathbf{F}_{i}^{\nabla p}$ are considered as the dominant interaction forces:

\begin{equation}
\mathbf{F}_{i}^{int}=\mathbf{F}_{i}^{D}+\mathbf{F}_{i}^{\nabla p}.\label{eq:Interaction}
\end{equation}

Various models are available for the calculation of the drag force.
In the current work, the correlation proposed by Di\,\,Felice {[}\citealp{DiFelice1994}{]},
which is well-anticipated in literature, is used:

\begin{equation}
\mathbf{F}_{i}^{D}=\frac{1}{8}C_{d}\rho_{f}\pi d_{i}^{2}(\mathbf{u}_{f,i}-\mathbf{v}_{i})|\mathbf{u}_{f,i}-\mathbf{v}_{i}|\varepsilon_{i}^{2-\chi},\label{eq:DiFelice}
\end{equation}
where $\varepsilon_{i}$, $d_{i}$, $\mathbf{u}_{f,i}$, $\mathbf{v}_{i}$
denote fluid fraction at solid particle $\mathcal{P}_{i}$ 
according Eq. \ref{eq:fluid-fraction} or Eq. \ref{eq:fluid-fraction-boundary} near boundaries,
the solid particle diameter, the fluid velocity and the solid particle
velocity, respectively. $\varepsilon_{i}$ is obtained from interpolating
fluid fractions of surrounding fluid particles $\mathcal{P}_a$:

\begin{equation}
\varepsilon_{i}=\frac{\underset{a}{\sum}\varepsilon_{a}V_{a}W_{ai}}{\underset{a}{\sum}V_{a}W_{ai}}
\end{equation}

The drag coefficient $C_{d}$ for spherical particles is given by:

\begin{equation}
C_{d}=\left(0.63+\frac{4.8}{\sqrt{Re_{i}}}\right)^{2}.\label{eq:DiFelice_DragCoef}
\end{equation}
where $Re_{i}=\varepsilon_{i}|\mathbf{u}_{f,i}-\mathbf{v}_{i}|d_{i}/\nu$
denotes the solid particle Reynolds number and $\nu$ denotes kinematic fluid viscosity.
$\chi$ is calculated as a function of the Reynolds number by

\begin{equation}
\chi=3.7-0.65\exp(-(1.5-\log_{10}(Re_{i}))^{2}/2).\label{eq:DeFelice_ksi}
\end{equation}

Provided that the pressure gradient $\nabla p$ arises only due to
interaction between solid particles and fluid, $F_{i}^{D}$
can be combined with $F_{i}^{\nabla p}$ {[}\citealp{Oschmann2014}{]}.
The latter results in:

\begin{equation}
\mathbf{F}_{i}^{int}=\frac{\mathbf{F}_{i}^{D}}{\varepsilon}-V_{i}\rho_{f}\mathbf{g}.
\end{equation}

$\mathbf{F}_{i}^{int}$ is used in Eq. (\ref{eq:sp_newton}) and Eq.
(\ref{eq:f_int_S-F}). For the mesh-based model, i.e. Eq. (\ref{eq:fvm_momentum}), on the other hand,
the force $\mathbf{f}_{m}^{int}$ acting on cell $c$ is calculated from the drag
forces $\mathbf{F}_{ci}^{D}$ which are acting in the cell:

\begin{equation}
\mathbf{f}_{k}^{int}=\underset{i}{\sum}\mathbf{F}_{ci}^{D}/V_{c}\:,
\end{equation}
where $V_{c}$ denotes the volume of the cell.

\section{Boundary conditions in SPH model\label{sec:Boundaries}}

\subsection{No-slip and No-penetration boundary model\label{sub:Boundary-model}}

The importance of accurately enforcing no-slip and no-penetration boundary conditions
(BC) was discussed in {[}\citealp{Adami2012,Marrone2013,Valizadeh2015}{]}.
Not only do BC models affect the accuracy of flow fields, but they 
also contribute to overall numerical stability {[}\citealp{Marrone2013}{]}.
Another aspect of boundary models in the context of the DEM-SPH method concerns
the convenience of use as pointed out in {[}\citealp{Sun2013}{]}.
In particular, it would be convenient if the geometry of the container, as 
defined for use in the DEM, could be directly used in SPH without further
effort. To satisfy the above requirements a new variant of the
BC model is proposed here. The new BC model allows container geometries
to be adopted directly from the DEM model. In comparison to the
BC model proposed in {[}\citealp{Sun2013}{]}, the BC model presented here
ensures no-slip conditions along the container walls.

\begin{figure}
\begin{centering}
\includegraphics[width=6cm]{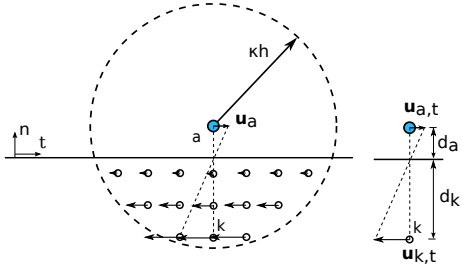}
\par\end{centering}
\caption{The calculation scheme of tangential boundary velocities \label{fig:BoundaryScheme}}
\end{figure}

A ghost-fluid technique is used to enforce no-slip and no-penetration boundary conditions.
The ghost-fluid technique is based on the idea of modeling
container walls using virtual fluid particles (ghost particles) positioned in the vicinity of
container walls. Every time step these ghost particles are instantaneously
generated for every fluid particle interacting with the boundary (see
Fig. \ref{fig:BoundaryScheme}). 
In contrast to the classical ghost
particle approach {[}\citealp{Colagrossi2003}{]} where fluid
particle is mirrored on the opposite side of the boundary line, in the proposed model,
several ghost particles are generated instantaneously to ensure that the support of the kernel interpolants is fully contained within the fluid phase.
A similar approach was used by
Marrone et al. {[}\citealp{Marrone2013}{]}, however, rather than instantaneously
generating ghost particles, they used pre-generated ghost particles. The instantaneous
generation of ghost particles is what distinguishes the proposed BC model from other
BC models proposed earlier.

As already pointed out earlier {[}\citealp{Adami2012,Eitzlmayr2014,Valizadeh2015}{]},
a special treatment of ghost particle velocities is needed to enforce no-slip boundaries
in a correct way. Following {[}\citealp{Marrone2013}{]}, two velocity fields are employed.
For the calculation of the viscous term in Eq. (\ref{eq:momentum-SPH}), the following ghost
particle velocities are used:

\begin{equation}
\begin{cases}
\mathbf{u}_{ak}^{(a)}\cdot\boldsymbol{\mathbf{t}}=\left[(\mathbf{u}_{a}-\mathbf{u}_{bc})\cdot(1+\nicefrac{d_{k}}{d_{a}})\right]\boldsymbol{\cdot\mathbf{t}},\\
\mathbf{u}_{ak}^{(a)}\cdot\mathbf{n}=(\mathbf{u}_{a}-\mathbf{u}_{bc})\cdot\mathbf{n},
\end{cases}\label{eq:bound-vel-visco}
\end{equation}
where $\mathbf{u}_{ak}^{(a)}$ denotes the velocity of the $k$-th ghost
particle relative to fluid particle $a$, $\mathbf{u}_{bc}$ denotes the prescribed 
velocity of the boundary, $\boldsymbol{\mathbf{t}}$ denotes a vector tangent 
to the boundary and $\mathbf{n}$ is a vector normal to the boundary. Furthermore,
$d_{k}$ and $d_{a}$ denote the normal distances of ghost particle $\mathcal{P}_{k}$
and fluid particle $\mathcal{P}_{a}$ from the container wall. For
the case of free-slip $\mathbf{u}_{ak}^{(a)}\cdot\mathbf{t}=0$
can be used. For the calculation of velocity difference in the continuity
equation Eq. (\ref{eq:continuity-SPH}), on the other hand, the following ghost
particle velocities are used:

\begin{equation}
\begin{cases}
\mathbf{u}_{ak}^{(a)}\cdot\mathbf{t}=0,\\
\mathbf{u}_{ak}^{(a)}\cdot\mathbf{n}=\left[(\mathbf{u}_{a}-\mathbf{u}_{bc})\cdot(1+\nicefrac{d_{k}}{d_{a}})\right]\cdot\mathbf{n}.
\end{cases}\label{eq:bound-vel-cont}
\end{equation}

Implementing two different ghost velocity fields avoids inconsistencies
and loss of accuracy as discussed by De Laffe et al. {[}\citealp{Leffe2011}{]}.
In particular, using Eq. (\ref{eq:bound-vel-cont}) in the continuity equation 
accounts for no-penetration whereas Eq. (\ref{eq:bound-vel-visco}) in the
momentum equation accounts for no-slip. Eitzlmayr et al. {[}\citealp{Eitzlmayr2014}{]} have
mentioned the problems of representing complex shaped geometries by discrete fluid particles and
suggest a way to avoid the generation of ghost particles by use of fitted polynomial functions.
In the presented BC model pre-generation of ghost particles was avoided by use of instantaneously
generated ghost particles. In contrast to the use of polynomial functions that represent boundary shapes,
the proposed BC model allows the use of arbitrary kernel functions without the need to adjust the BC model.

\subsection{Validation tests}

Two 2D tests, namely the Poiseuille flow and the flow through a periodic
lattice of cylinders, are performed to validate the proposed BC model.
Such or similar tests are used by many researchers to validate no-slip
BCs in the SPH.

\subsubsection{Poiseuille flow}

The test case with two infinite parallel walls and fluid in between
(Poiseuille flow) is used to verify the described boundary conditions.
The fluid particles are initially at rest and driven by a body force
applied in the horizontal direction. The 2D flow with 19 fluid particles
in horizontal direction and 25 fluid particles in vertical direction is
considered. The simulation parameters as used by Eitzlmayr et al.
{[}\citealp{Eitzlmayr2014}{]} are used: the smoothing length
is $h=0.24\,\mathrm{mm}$, initial distance between fluid particles is 0.2\,mm,
the fluid density is $\rho=1000\,\mathrm{kg/m^{3}}$, the fluid viscosity
is $\mu=0.5\,\mathrm{Pa\cdot s}$, the speed of sound (see. Eq. (\ref{eq:state}))
is $c=10\,\mathrm{m/s}$, the body force is $10\,\mathrm{m/s^{2}}$.
The analytical solution for Poiseuille flow can be found in {[}\citealp{Morris1997,Eitzlmayr2014}{]}. 

\begin{figure}
\begin{centering}
\includegraphics[width=10cm]{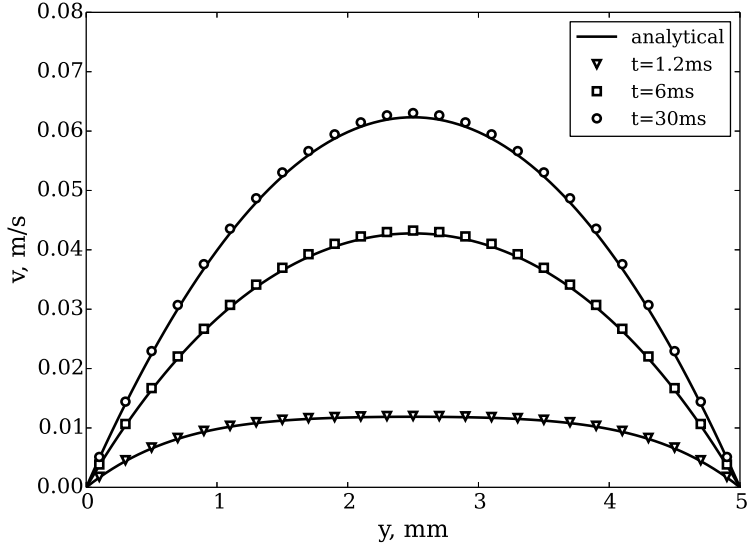}
\par\end{centering}

\caption{Poiseuille flow\label{fig:pipe_velocities}}
\end{figure}

The velocity profiles for the analytical and numerical SPH solutions for 1.2,
6 and 30\,ms after applying the body force are presented in Fig.
\ref{fig:pipe_velocities}. In general, there is good agreement between
the obtained results. The SPH velocities are slightly larger, than
the analytical, but results correspond to the obtained values in {[}\citealp{Morris1997}{]}
and in {[}\citealp{Eitzlmayr2014}{]}. The velocities are approaching
zero values near to the boundaries, which indicates that the no-slip
condition is enforced correctly. 

\begin{figure}
\begin{centering}
\includegraphics[width=4cm]{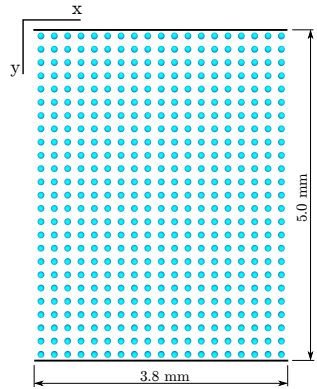}
\par\end{centering}

\caption{Initial fluid particle setup for Poiseuille flow\label{fig:pipe_particles}}
\end{figure}

\subsubsection{Flow through a periodic lattice of cylinders}

\begin{figure}
\begin{centering}
\includegraphics[width=10cm]{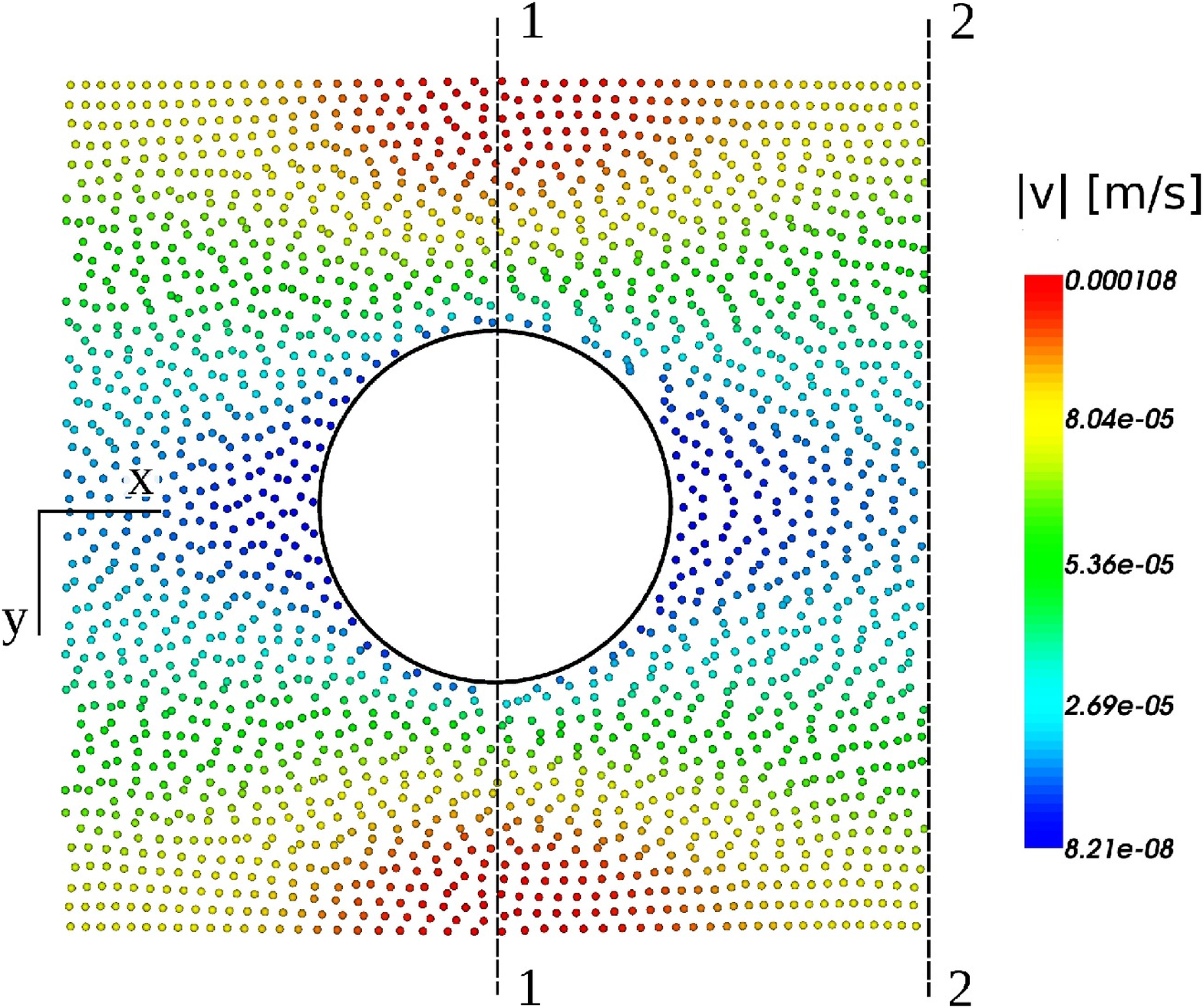}
\par\end{centering}

\caption{Fluid particle positions at t=4000\,s colored according to velocity magnitudes
\label{fig:cil_Particle-positions}}
\end{figure}

Another test case used for validation of the presented boundary conditions
is the 2D flow through a square lattice of cylinders. A single cylinder
with an associated volume within the lattice is considered in the
SPH (Fig. \ref{fig:cil_Particle-positions}). As in the previous test
case, the flow is driven by a body force. The periodic boundaries are
applied in x and y directions, while the no-slip boundary conditions
presented in section \ref{sub:Boundary-model} are applied for the
cylinder. Although the used boundary model is easy to adapt to cylindrical
surfaces (simply, ghost particles could be generated below the cylinder
surface following its curvature), such adaptation was not used here.
The ghost particles were generated assuming a plain surface of the
boundary. This simplification is reasonable in this case, because
the diameter of the used cylinder is much larger than the distance
between the SPH particles.

\begin{figure}

\begin{centering}
\includegraphics[width=8.5cm]{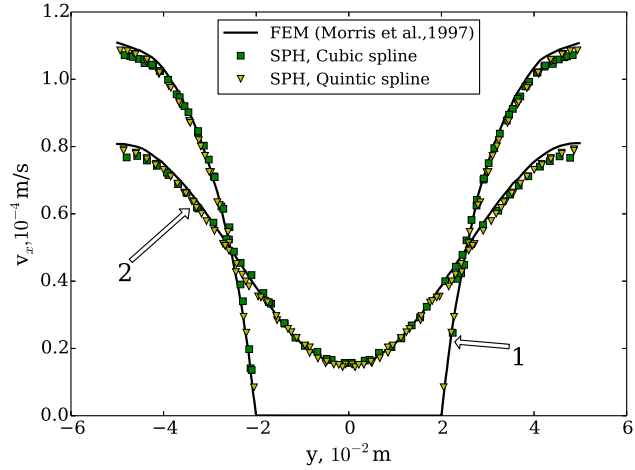}
\par\end{centering}

\caption{Velocity profile along cut lines shown in Fig. \ref{fig:cil_Particle-positions}
for flow through a periodic lattice of cylinders \label{fig:cil_velocities}}

\end{figure}

The parameters for the test case are the same as used by Morris et
al. {[}\citealp{Morris1997}{]}: the size of the domain is 0.1x0.1\,m,
the initial distance between the fluid particles is 2\,mm, the viscosity
is $\mu=10^{-3}\,\mathrm{Pa\cdot s}$, the smoothing length is $h=2.4\,\mathrm{mm}$,
the fluid density is $\rho_{f}=1000\,\mathrm{kg/m^{3}}$, the speed
of sound is $c=5.77\cdot10^{-4}\,\mathrm{m/s}$, the body force is
$1.5\cdot10^{-7}\,\mathrm{m/s^{2}}$, the diameter of the cylinder
is 4\,cm. The background pressure $B=10^{-5}\,\mathrm{Pa}$ (see
Eq. \ref{eq:state}) is used to avoid the negative pressure in the
downstream flow and the formation of an unphysical void formation as a
consequence. Initially the velocities of the fluid particles are zero. The
fluid particles start to move due to the initiation of the body force. The
velocities increase until steady state is reached. Simulations using
a cubic spline (Eq. (\ref{eq:KernelSpline})) and a quintic spline
kernel {[}\citealp{Morris1997}{]} are performed. The fluid particles at
the final time $t=4000\,\mathrm{s}$ colored by the velocities are
shown in Fig. \ref{fig:cil_Particle-positions}. The velocity profiles
along the cut lines 1 and 2 from SPH (Fig. \ref{fig:cil_Particle-positions})
together with the results from the steady incompressible viscous flow
using a Finite Element Method (FEM) {[}\citealp{Morris1997}{]} are
presented in Fig. \ref{fig:cil_velocities}.

The results using both kernels are close to the results obtained by
FEM. The velocities with the quintic spline kernel are a little bit
closer to the results obtained by FEM. This corresponds to the discussion
presented in {[}\citealp{Morris1997}{]}. However because the quintic
spline is computationally more expensive, the simpler cubic spline
kernel is further used in the current research.

\section{Solid particle sedimentation test\label{sec:Single-part-test}}

\begin{figure}
\begin{centering}
\includegraphics[width=5cm]{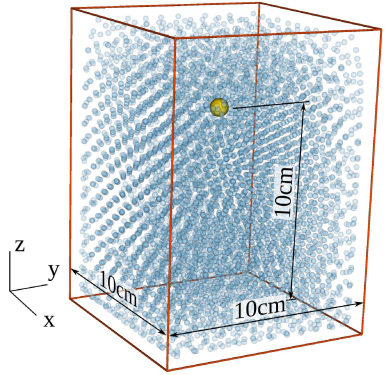}
\par\end{centering}

\caption{Solid particle inside the container filled with SPH particles\label{fig:sp_initial_positions_sph}}

\end{figure}

To examine the performance of DEM-SPH and DEM-FVM methods, three numerical
tests, starting from the settlement of one solid particle, are performed.
In this single particle sedimentation test a solid spherical particle
is placed in a 3D container with fluid and is realized to fall down
under the influence of the gravity force. The density $\rho_{f}=1000\,\mathrm{kg/m^{3}}$
and the viscosity $\mu=0.001\,\mathrm{Pa\cdot s}$ are used for the
fluid in both SPH and FVM methods. These physical fluid parameters
are used in all following test cases. The density of the material
of the solid particle is set to $\rho_{s}=1200\,\mathrm{kg/m^{3}}$.
Three solid particle diameters $d=2,\,4,\,8\,\mathrm{mm}$ are considered
in the tests. SPH particles with a smoothing length $h=8\,\mathrm{mm}$
and an initial distance $\triangle x=5.33\,\mathrm{mm}$ between them
are generated in the container above the bottom wall. This gives $h/\triangle x=1.5$,
which ensures that enough neighbors are around every fluid particle
{[}\citealp{Monaghan2005}{]}. In total 8000 SPH particles are used.
As an initial preparatory step, the fluid particles are allowed to
settle in the container. Solid and fluid particle positions after this preparatory
step are shown in Fig. \ref{fig:sp_initial_positions_sph}. The boundary
conditions described in section (\ref{sub:Boundary-model}) are used
for the walls of the container. 

To be able to compare the DEM-SPH and DEM-FVM, both methods should
be expected to give the same (similar) results. The drag force acting
onto the solid particles highly depends on the fluid fraction (see
Eq. (\ref{eq:DiFelice})). Therefore, the size of the cell in the
FVM is chosen to give the same fluid fraction as in the SPH. With
the current DEM-SPH setup and when the solid particle with $d=8\,\mathrm{mm}$
is used, the calculated fluid fraction during the simulation is approximately
$\varepsilon=0.935$. This gives us the cell size $0.0143\times0.0143\times0.0202$,
which is used for the DEM-FVM. As a result, the container is divided
into $7\times7\times6$ cells.

\begin{figure}
\begin{centering}
\includegraphics[width=10cm]{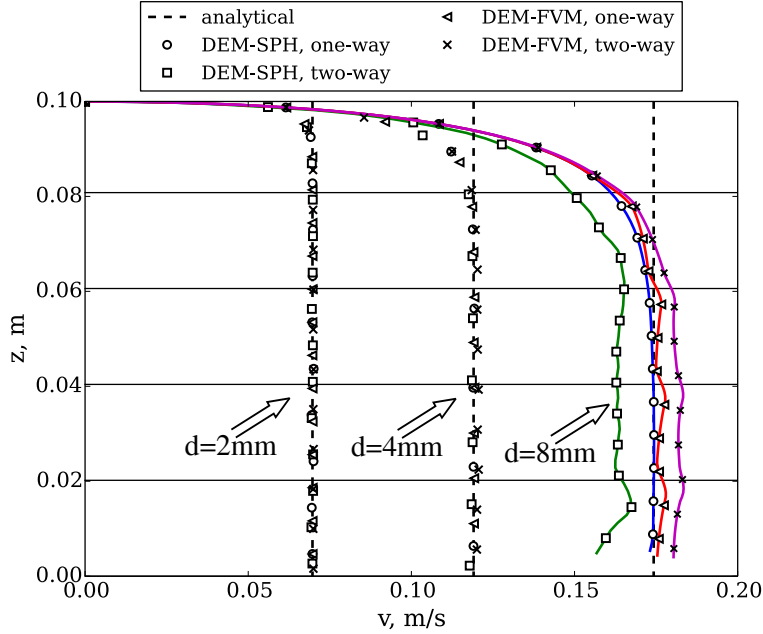}
\par\end{centering}

\caption{Solid particle sedimentation velocity \label{fig:Particle-sedimentation-velocity}}
\end{figure}

The velocities of the settling solid particles are shown in Fig.\ref{fig:Particle-sedimentation-velocity}.
The curves present results obtained using a one-way coupling and a
two-way (full) coupling between DEM and SPH or FVM. In the one-way
coupling scheme, the solid particles are experiencing the presence
of the fluid, however the fluid does not ``feel'' the presence of
the solid particles, i.e. the fluid fraction of fluid remains one
and the source term on the fluid is always zero in the FVM (see Eq.
\ref{eq:fvm_momentum}) and in the SPH (see Eq. \ref{eq:momentum}).
Together with the numerical results, analytical terminal velocities
of the solid particles are presented in the figure as dashed vertical lines.
The analytical terminal velocity was calculated from the sum of the
gravity, drag and bouyancy forces, which should be equal to zero when
terminal velocity is reached:

\begin{equation}
m\mathbf{g}+\mathbf{F}^{D}-V\rho_{f}\mathbf{g}=0,\label{eq:vel_terminal_calc}
\end{equation}
where $V$ is the volume of the solid particle. The drag force $\mathbf{F}^{D}$
was calculated using the same drag force correlation (Eq. (\ref{eq:DiFelice}))
as was used in the numerical methods, using the same constant fluid
fraction $\varepsilon=0.999$ for $d=2\mathrm{mm}$, $\varepsilon=0.992$
for $d=4\mathrm{mm}$, $\varepsilon=0.935$ for $d=8\mathrm{mm}$
as was obtained from the DEM-SPH and the fluid velocity $\mathbf{u}_{f}$
equal to zero. With this assumptions for the drag force, the analytical
solution corresponds to the one-way coupling scheme in the numerical
tests. Horizontal lines in Fig. \ref{fig:Particle-sedimentation-velocity}
show the cell boundaries in the FVM.

In the tests with the solid particle $d=2\mathrm{mm}$ there are almost
no difference between all 5 results (analytical, DEM-SPH one-way,
DEM-SPH two-way, DEM-FVM one-way, DEM-FVM two-way). In the tests with
the solid particle of $d=4\mathrm{mm}$ very small differences can be recognized.
However the tests with a solid particle of $d=8\mathrm{mm}$ show differences
between the calculated solid particle velocities. The obtained velocity
in the DEM-SPH one-way coupling fully overlaps the line of the analytical
terminal velocity. The DEM-FVM one-way result shows a bit higher solid particle
settlement velocity. It is related to the way the fluid fraction is
calculated on the solid particle. When the solid particle crosses a cell
boundary, a part of the particle volume is assigned to one cell, while
another part is assigned to another cell and, accordingly, the resultant
fluid fractions are higher. Only when the particle is fully enclosed
in one cell, the calculated fluid fraction corresponds to the value
used in the analytical solution. This change of the fluid fraction
is reflected in the waving character of DEM-FVM curve. There is interesting
difference obtained between the results using two-way coupling. Because
of the source term applied to the fluid in the FVM, the velocity vector
in the cell, where the solid particle is, is pointing downwards. This results
in a bit smaller velocity difference between the velocity of the particle
and the fluid. Consequently, a smaller drag force is obtained and
the particle moves a bit more quickly. 

The opposite picture is obtained with the DEM-SPH. Here some fluid
particles near the centre of the solid particle move in the opposite
direction then the solid particle and, therefore, a bit bigger velocity
difference is obtained. Consequently, the bigger drag force is obtained
and the solid particle moves slower. This result corresponds to the
results reported in {[}\citealp{Robinson2014}{]}. The authors in
{[}\citealp{Robinson2014}{]} considered settlement of a single solid particle
using different fluid resolution ranging $h/d$ from 1.5 to 6. In
comparison with $2\leq h/d\leq6$, the lower settlement velocity was
obtained in the case of $h/d=1.5$. Because in our test case $h/d=8\,\mathrm{mm}/8\,\mathrm{mm}=1$,
the same trend should be expected.

It should be noted, that strictly speaking the application of the
local averaging technique (Eqs.(\ref{eq:continuity})-(\ref{eq:momentum}))
for the prediction of the movement of a single solid particle is incorrect.
However we used it as a test case to clarify possible differences
between the DEM-SPH and DEM-FVM methods. The test cases shown in the
next sections deal with the assembly of solid particles, i.e. the
case the averaging technique is developed for.

\section{Flow through a porous block\label{sec:gravity-block-test}}

\begin{figure}
\begin{centering}
\includegraphics{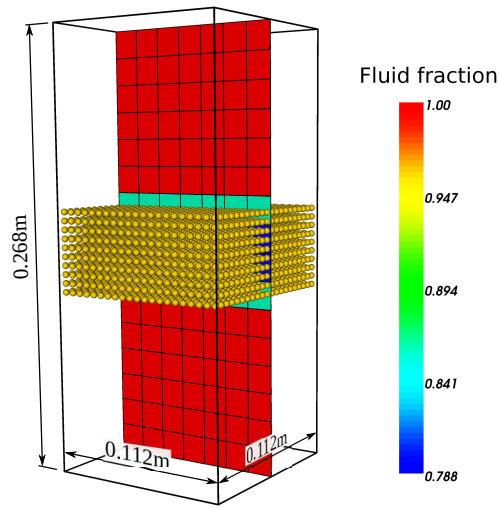}
\par\end{centering}

\begin{centering}
\caption{Container with the porous block. The cross section of the cells is
coloured by the fluid fraction as obtained in the DEM-FVM \label{fig:gb-Container-scheme}}

\par\end{centering}

\end{figure}

A numerical analysis of fluid penetrating through a fixed porous block
is performed. The porous block is constructed from solid particles
fixed in space. The diameter of the solid particles is $d=4\,\mathrm{mm}$,
while the distance between the solid particle centers is $5.33\,\mathrm{mm}$.
This gives the fluid fraction inside of the block as $\varepsilon=0.779$.
$21\cdot21\cdot10$ solid particles are used to resemble the porous
block. 21000 SPH particles with a smoothing length of $h=8\,\mathrm{mm}$
and an initial distance $\triangle x=5.33\,\mathrm{mm}$ between them
are generated in the container. The container is divided into $7\cdot7\cdot16$
cells for analysis with the DEM-FVM. The block is placed in a container
with fluid (see Fig. \ref{fig:gb-Container-scheme}) and remains fixed during the simulation. The density
$\rho_{f}=1000\,\mathrm{kg/m^{3}}$ and the viscosity $\mu=0.001\,\mathrm{Pa\cdot s}$
are used for the fluid. Free slip wall boundary conditions are used
for sides of the container. Periodic boundaries are applied on the
top and the bottom of the domain. At first, an initialization simulation
is performed to reach an equilibrium state. Then, the gravity $9.81\,\mathrm{m}/\mathrm{s}^{2}$
is applied to the fluid and simulations using both numerical methods
(DEM-SPH and DEM-FVM) are performed. The fluid, which has zero velocity
initially, accelerates until a constant flow velocity is reached.

\begin{figure}
\begin{centering}
\includegraphics[width=10cm]{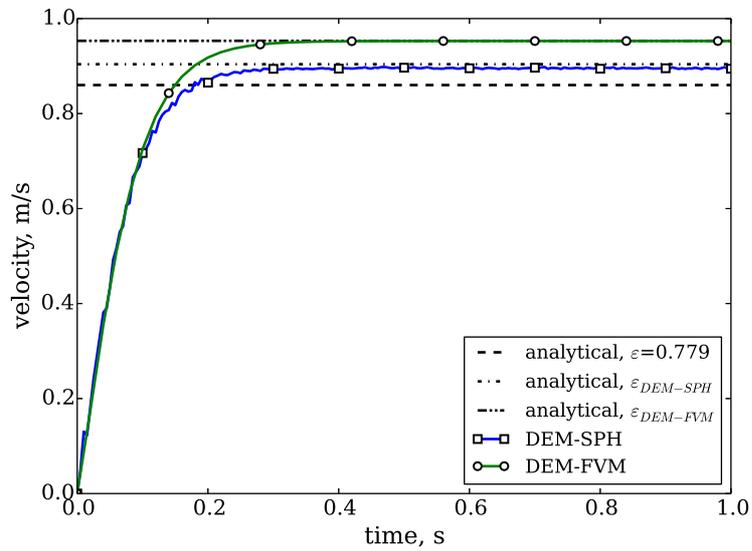}
\par\end{centering}

\caption{Fluid flow velocities \label{fig:gb-Flow-velocity}}
\end{figure}

Velocities obtained by the simulations together with analytically calculated velocities
are shown in Fig. \ref{fig:gb-Flow-velocity}. Fluid velocities shown in 
Fig. \ref{fig:gb-Flow-velocity} from FVM and SPH results were taken as average
velocities on a cross plane $x'O'y'$ shown in Fig. \ref{fig:gb-Container-scheme}. Analytical velocities
are calculated from Eq. (\ref{eq:vel_terminal_calc}), however, this
time, the mass $m$ is the total mass of the fluid, while $\mathbf{F}^{D}$ and $V$ are the
total drag force and the total volume of the solid particles respectively.

\begin{figure}

\begin{centering}
\includegraphics[width=10cm]{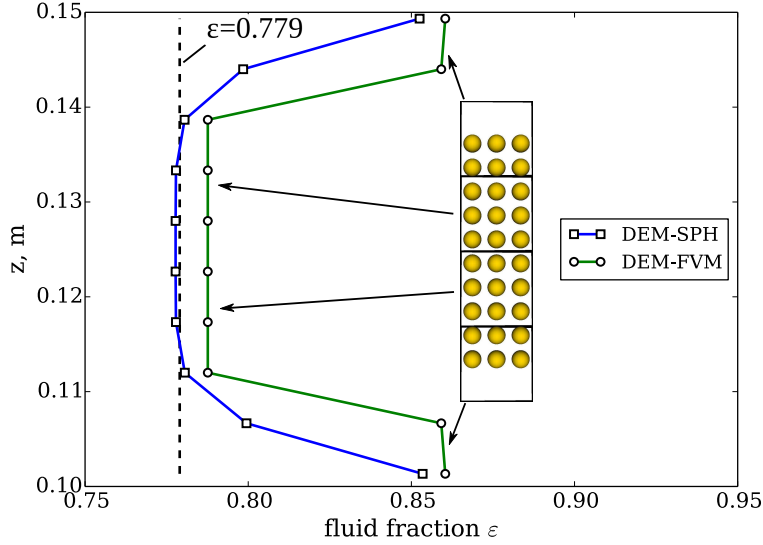}
\par\end{centering}

\caption{Fluid fraction inside the porous block in the DEM-SPH and the DEM-FVM\label{fig:gb-ffraction-inside-block}}

\end{figure}

While the same porous
block was used in both numerical methods, the resulting constant velocity
for the DEM-FVM is $0.953\,\mathrm{m/s}$, but for the DEM-SPH it
is $0.896\,\mathrm{m/s}$. Both results are higher than the analytically
calculated value of $0.856\,\mathrm{m/s}$ for $\varepsilon=0.779$.
The difference is the result of the different fluid fractions at the
positions of the solid particles. As it is shown in Fig. \ref{fig:gb-ffraction-inside-block},
solid particle layers of the block have a bit different fluid fractions.
These differences are causing the different drag forces calculated
in two methods, because the calculated drag force highly depends on
the fluid fraction (see Eq. (\ref{eq:DiFelice})). The fluid velocity
(Fig. \ref{fig:gb-Flow-velocity}) obtained with the DEM-FVM fully
overlaps the corresponding analytical line. However small differences
can be seen when comparing the DEM-SPH result with the corresponding
analytical line. This difference (no more than 1\%) could be caused
by the influence of the walls. When the fluid fraction is less than
one (situation inside of the porous block), the truncated kernels
are taken into account by the use of the $\Gamma$ factor in Eq. (\ref{eq:fluid-fraction-boundary}).
Still, by the use of the $\Gamma$ factor the influence of the walls
is estimated only approximately. The inaccuracy made by the use of
Eq. (\ref{eq:fluid-fraction-boundary}) decrease with the use of smaller
solid particles. Additionally, the use of the $\Gamma$ factor can
not take into account the situation on the corners where two walls
are intersecting. The result of this is a bit smaller fluid fraction
near the sides of the container where the porous block is present.

\section{Sedimentation of a porous block\label{sec:Porous-block-test}}

\subsection{Initialization of the specimen}

\begin{figure}
\begin{centering}
\includegraphics[width=7cm]{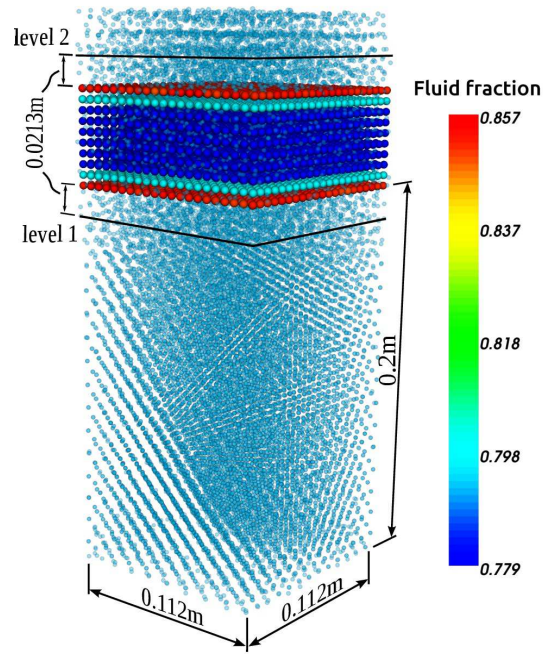}
\par\end{centering}

\caption{Container with the porous block (DEM-SPH) after initialisation\label{fig:pb_container_init_sph}}

\end{figure}

Sedimentation of a constant porosity block under gravity is considered
in this test. Both DEM-SPH and DEM-FVM simulations are performed.
As in the previous test case, the porous block is formed from $21\cdot21\cdot10=4410$
solid particles with a density of $1200\,\mathrm{kg/m^{3}}$, which
have the diameter $d=0.004\mathrm{\, m}$, and are placed with $0.00533\,\mathrm{m}$
distance between their centers. The same size of container and the same number
of cells for DEM-FVM and the same number of fluid particles for DEM-SPH
as in the previous test case are used. However, in this test a wall
is defined at the bottom, while periodic boundaries are used on the sides
of the container. Initially the generated porous
block is placed 20 cm above the bottom wall (Fig. \ref{fig:pb_container_init_sph}). 
After generating the
solid particles and the fluid cells, a steady state simulation is
performed to reach an equilibrium state by DEM-FVM. In the DEM-SPH, after the
generation of solid particles,
the fluid particles are generated in the container. During
the generation, the distances between the fluid particles are adjusted
to take into account the lower fluid fraction inside the porous block.
Then initial simulation of the system is performed by keeping the
porous block unmoved. At first stage, some movement of the fluid particles
are produced, because the fluid particles try to find equilibrium positions.
As a result, the variation of the fluid pressure is produced.
However after about 1\,s, the almost constant value of the pressure
is reached. It should be noted, that some level of chaotic movement
of the fluid particles remains. However such movements are common
for SPH {[}\citealp{Sivanesapillai2015}{]}. 

\begin{figure}
\begin{centering}
\includegraphics[width=10cm]{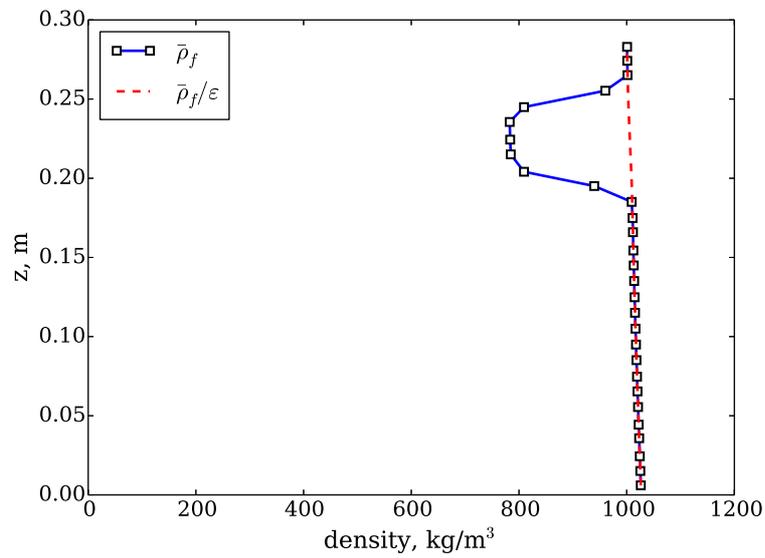}
\par\end{centering}

\caption{Fluid density distribution at the end of the initialization procedure in the
DEM-SPH\label{fig:pb_densities}}
\end{figure}

After the initial simulation the resultant density distribution inside
of the container simulated by the DEM-SPH is shown in Fig.\ref{fig:pb_densities}.
The superficial density $\bar{\rho_{f}}$ inside of the porous block
is reduced because solid particles occupy volume and the fluid particles
are forced to moved out from the block. The value of the superficial
density is equal to $\bar{\rho_{f}}=\rho_{f}\cdot\varepsilon$=$1000\cdot0.779$=
$\mathrm{779\, kg/m^{3}}$ inside the center of the block. However
near to the top and bottom of the block the transition domain can
be seen which reflects the smooth change of the fluid fraction near
the boundaries of the block. If the superficial density is divided
by the fluid fraction, the physical fluid density is obtained (second
line in Fig. \ref{fig:pb_densities}). In Fig. \ref{fig:pb_densities}
some increase of the calculated fluid density with the depth can be
seen. It is because of the weakly compressible approach used in the
SPH (see Eq.(\ref{eq:state})).

\begin{figure}

\begin{centering}
\includegraphics[width=10cm]{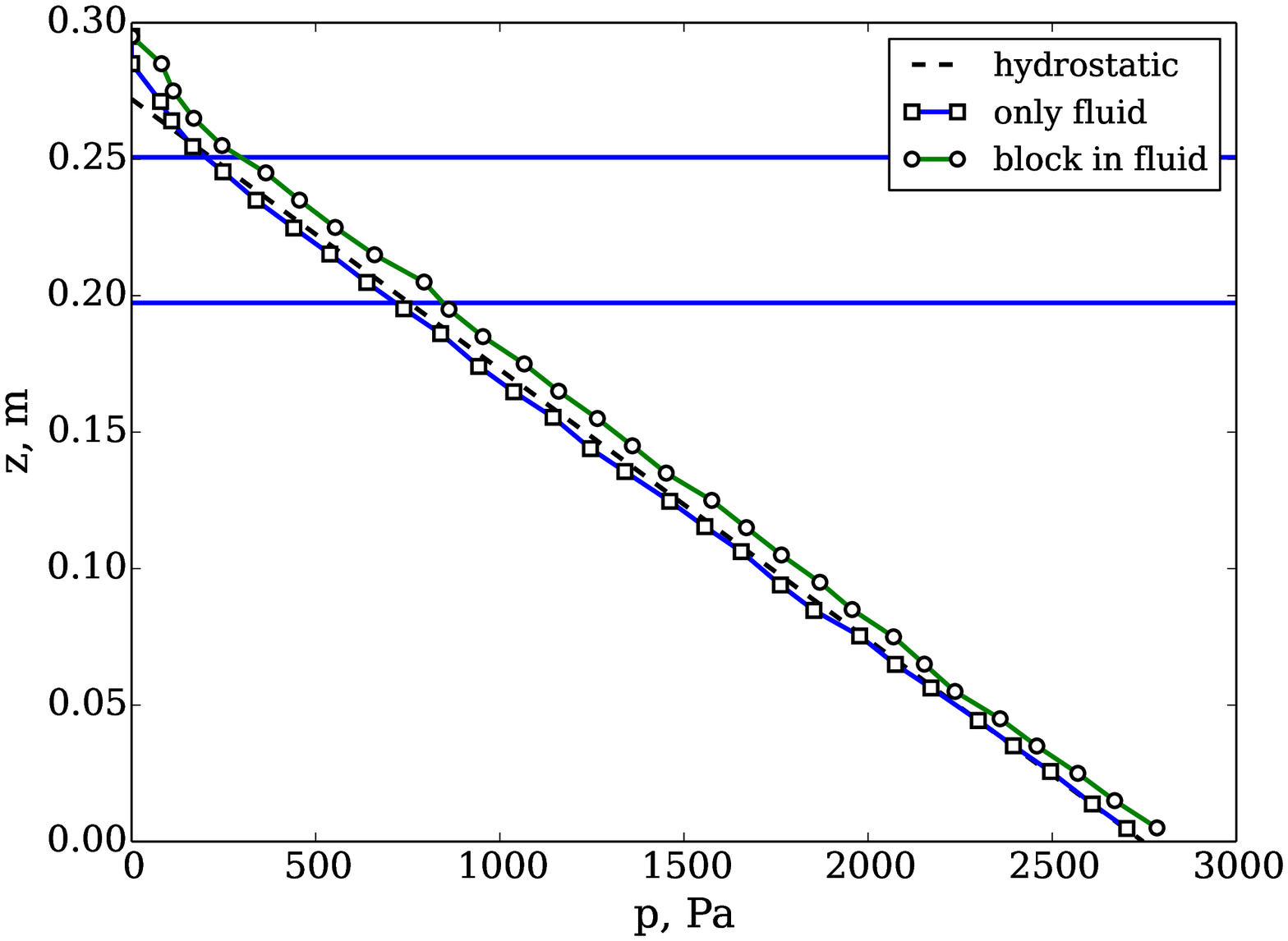}
\par\end{centering}

\caption{Pressure in the container at the end of the initialisation procedure in the
DEM-SPH \label{fig:pb_press_vs_z}}
\end{figure}

The calculated pressure in the container (Fig. \ref{fig:pb_press_vs_z})
increases from zero at the free surface to $2780\,\mathrm{Pa}$ near
the bottom (line ``block in fluid''). Additionally to the case of
the container with the porous block inside (line ``block in fluid'')
there was performed a simulation with the fluid only (line ``only
fluid''). As is expected the pressure is higher in case when the
block is submerged in the fluid, because of the increased fluid level.
However the expected difference is 115$\,$Pa, while the calculated
difference at the bottom is equal to 82$\,$Pa only. At higher positions
this difference is higher. An analytically obtained hydrostatic line
($p=\rho_{f}g\Delta z$) is almost overlapped by the ``only fluid''
line, while the ``block in fluid'' line keeps above however more
or less parallel to it. Two horizontal lines in Fig.\ref{fig:pb_press_vs_z}
show the top and bottom positions of the block. Bigger discrepancies
from the hydrostatic line can be seen near the free surface, because
of the truncated support domain of the kernel near the free surface
{[}\citealp{Lee2008}{]}.

\subsection{Sedimentation step}

After initialization of the specimen, the solid particles are released
and they start to settle down by the influence of the gravity force.
To reduce pseudo-sound waves in the fluid domain, the gravity force
$F_{g}$ on the solid particles is increased gradually by using Eqs.(\ref{eq:f_g_damping})-(\ref{eq:t_damp})
with $t_{damp}=0.5\,\mathrm{s}$. Relative movements of the solid
particles in the block are not allowed, therefore the particles settle
down as a one solid block.

\begin{figure}
\begin{centering}
\includegraphics[width=10cm]{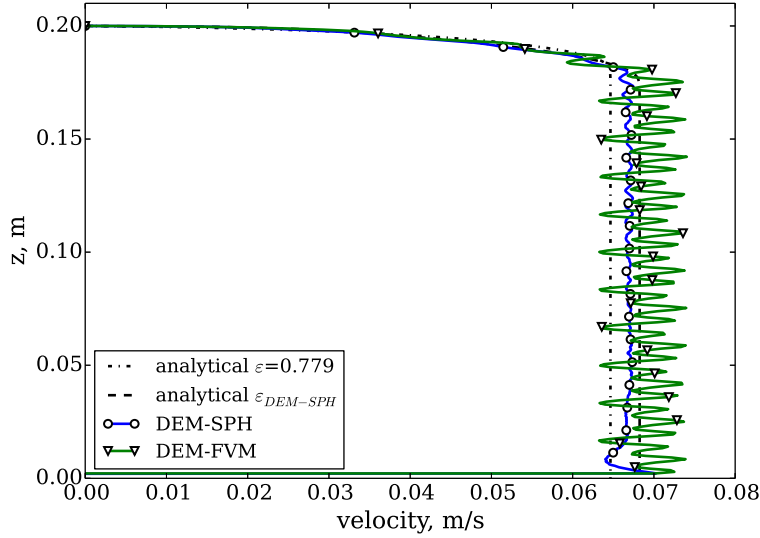}
\par\end{centering}

\caption{Settling velocity of the porous block \label{fig:pb_velocity}}
\end{figure}

The result of the simulation of the block sedimentation by means of
a settlement velocity is presented in Fig.\ref{fig:pb_velocity}.
Together with the DEM-SPH and DEM-FVM results, analytically calculated
curves are drawn. The analytical curves are obtained by numerically
integrating the solid particle acceleration from the out-of-balance force
$\mathbf{F}^{oob}$ over the time:

\begin{equation}
\mathbf{F}^{oob}=\mathbf{F}^{g}+\mathbf{F}^{D}+\mathbf{F}^{b}.
\end{equation}
The $\mathbf{F}^{D}$ is calculated using Eq.(\ref{eq:DiFelice})
with $(\mathbf{u}_{f,i}-\mathbf{v}_{i})=-\mathbf{v}_{i}/\varepsilon$,
because all the fluid in the container is forced to penetrate through
the porous block during its settlement. The first analytical curve
is obtained by assuming a constant fluid fraction $\varepsilon=0.779$
inside the whole block, while the second analytical curve is obtained
using the fluid fraction taken from the DEM-SPH, which has different
values in solid particle layers near to the block boundaries. In the DEM-SPH
simulation the block reaches its maximum velocity after about 0.5\,s
(the same time as $t_{damp}$, $z=0.175\,\mathrm{m}$) and moves down
until it reaches the bottom wall at $t=3.14\,\mathrm{s}$. While the
applied damping technique reduced the pseudo-sound waves, some waving
can be seen on the block velocity curve. The influence of the bottom
boundary is reflected in the last part of the curve ($z=0-0.015\,\mathrm{m}$). 

\begin{figure}
\begin{centering}
\includegraphics[width=7cm]{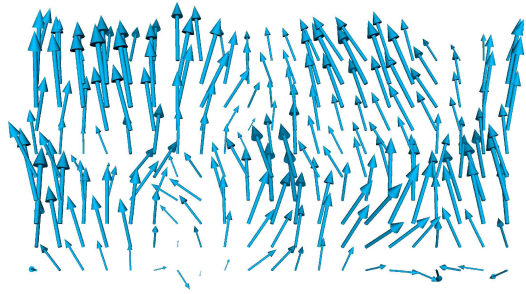}
\par\end{centering}

\caption{Fluid velocities at positions of the solid particles at time $t=2.3\,\mathrm{s}$,
$|\mathbf{u}_{f}|_{max}=0.0435\,\mathrm{m/s}$ (DEM-SPH) \label{fig:pb_fvelocity_vector}}
\end{figure}

There is a difference about 2\% between the settlement velocity obtained
from DEM-SPH simulation and the analytical curve even when the fluid
fractions from DEM-SPH are used to obtain velocity. This difference
is caused by inhomogeneous distribution of the velocities at the positions
of solid particles (see Fig. \ref{fig:pb_fvelocity_vector}). For
example at $t=2.3\mathrm{\, s}$ the drag force calculated using one
velocity for the whole porous block is equal about 0.251\,N, while
the drag force calculated in DEM-SPH (where the drag force is calculated
using the individual velocity values at each position of the solid
particle) is $F_{DEM-SPH}^{d}$ =0.262\,N. For the whole porous block
$(F^{g}+F^{b})=0.263\,\mathrm{N}$. The small difference between $F_{DEM-SPH}^{d}$
and $(F^{g}+F^{b})$ equal to 0.001\,N at this time step causes the
small increase of velocity in the next time step.

\begin{figure}
\begin{centering}
\includegraphics[width=10cm]{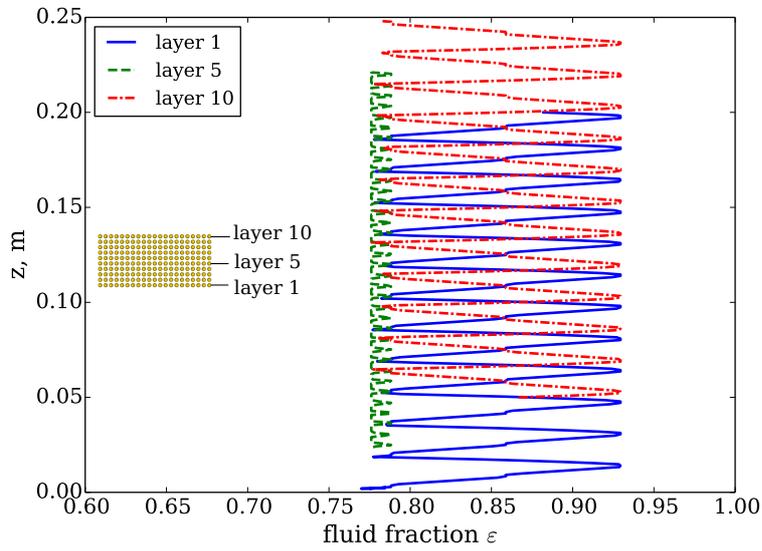}
\par\end{centering}

\caption{The variation of the fluid fraction during the sedimentation of the
porous block in the DEM-FVM\label{fig:pb_ffraction_variation}}

\end{figure}

The velocity curve of the DEM-FVM simulation shows that the block
does not settle at a constant velocity, however the velocity is changing
periodically. This change of the velocity is related to the change
of the fluid fraction on the positions of the solid particles. The
variation of the fluid fraction in the layers of the solid particles
is presented in Fig.\ref{fig:pb_ffraction_variation}. When the block
moves down, the calculated fluid fraction is changing because in the
used algorithm for calculating the fluid fraction, when a solid particle
crosses the boundary between two cells, its volume is divided proportionally.
The fluid fraction varies a lot especially at the upper and bottom
layers of the block ($0.778-0.929$), while in the middle the variation
is between 0.776 and 0.788 only. The result of such a change of the
fluid fraction is that a different total drag force is calculated.
The change of the calculated fluid fraction could be reduced, if finer
cells for the FVM would be used. In contrast to DEM-FVM, in DEM-SPH
the fluid fraction at the positions of the solid particles is varying
much less during the settlement of the block. Therefore the block
in DEM-SPH moves down at a more constant velocity.

\begin{figure}
\begin{centering}
\includegraphics[width=10cm]{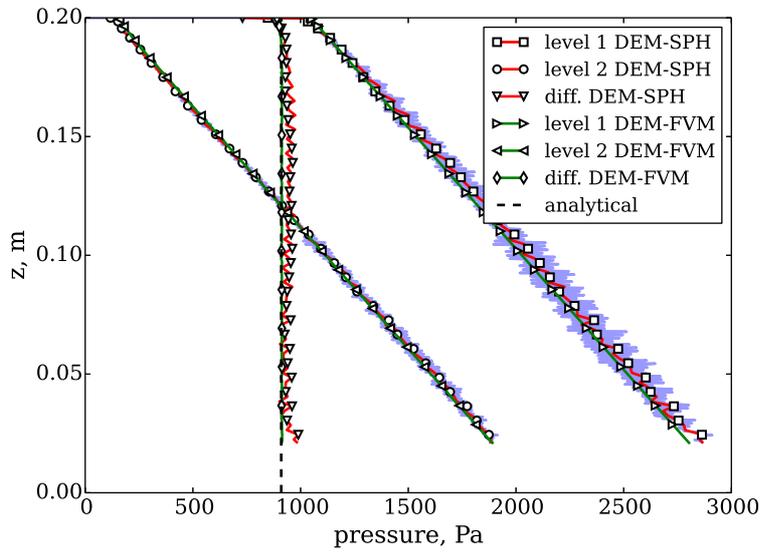}
\par\end{centering}

\caption{Pressures during the sedimentation of the block\label{fig:pb_press_vs_time}}
\end{figure}

The pressures in the fluid below (level\,1, Fig. \ref{fig:pb_container_init_sph})
and above (level\,2, Fig. \ref{fig:pb_container_init_sph}) the porous
block during the settlement are presented in Fig. \ref{fig:pb_press_vs_time}.
The analytical vertical line shows expected difference as is calculated
from

\[
\Delta p=\rho_{f}|\mathbf{g}|\Delta z+\frac{|\mathbf{F}^{g}|-|\mathbf{F}^{b}|}{A_{s}},
\]
where the first term on the right side is the pressure difference
due to difference in the hydrostatic pressure at level 1 and level
2. The second term reflects the pressure difference due to the drag
force, which, in the case when the terminal velocity is reached, is
equal to the sum of the gravity and bouyancy forces. $A_{s}$ is the
area of the cross section of the container. The ``diff. DEM-SPH''
curve shows the difference between pressures at level\,1 and level\,2.
Quite big fluctuations can be seen in the pressures. However the difference
between the expected analytical value and the averaged numerical value
is about 5\,\% only. In contrast to DEM-SPH results, there are no
fluctuations in DEM-FVM results. The resulting pressure difference
(``diff. DEM-FVM'' curve in Fig. \ref{fig:pb_press_vs_time}) overlaps
the analytical calculated line. It seems that the SPH method has difficulties
to handle the pressure field in this case. The fluctuating pressure
problem in the SPH is also reported by other researchers {[}\citealp{Molteni2009,Antuono2010}{]}.

\section{Conclusions}

In the present work, a comparative study on mesh-based and mesh-less coupled CFD-DEM
methods to model particle-laden flow was performed. The governing equations describing
the coupling of the Discrete Element Method with the Smoothed Particle
Hydrodynamics method were presented in detail. Comparative DEM-FVM and DEM-SPH
tests were performed and similarities and differences were
discussed. Based on this work, the following conclusions
can be drawn:
\begin{itemize}
\item The proposed new model to account for boundary conditions in the DEM-SPH approach
was demonstrated to produce accurate results in the presented verification tests.
They proved to be convenient and stable in the context of our DEM-SPH simulations.
\item In general, results obtained using DEM-FVM and DEM-SPH approaches agreed well
with analytic reference results. Numerical difference between DEM-SPH and DEM-FVM
were found mostly due to difference in computed fluid fractions that result in
different drag forces. 
\item In some cases, e.g. in the porous block settlement test, the DEM-FVM
shows an unsmooth settlement velocity curve. This is caused by the
constantly changing fluid fraction when solid particles are mapped
from one cell to another. The settlement curve obtained with DEM-SPH
remains smooth.
\item Due to weak compressibility of the present SPH scheme, pressure fluctuations are 
observed during the settlement of the porous block in the DEM-SPH approach. This corresponds
to the results reported by Robinson et al. {[}\citealp{Robinson2014}{]}, where an additional
artificial viscosity was used to dampen these fluctuations. However,
even without the artificial viscosity the mean values of fluid pressures
reproduce analytical reference results with satisfactory accuracy.
\end{itemize}

\section*{Acknowledgements}

This project has received funding from the European Union\textquoteright{}s
Horizon 2020 research and innovation programme under the Marie Sklodowska-Curie
grant agreement No 652862.

\section*{References}

\bibliographystyle{elsarticle-num}
\addcontentsline{toc}{section}{\refname}\bibliography{sph-dem_15.bib}

\end{document}